\documentclass[cits,proceedings,nohyper]{PoS}

\usepackage{amsmath,amssymb,amsfonts}
\usepackage{amsthm}
\usepackage[english]{babel}
\usepackage[latin1]{inputenc}
\usepackage{fontenc}
\usepackage{times}
\usepackage{mathptmx}
\usepackage{graphicx}
\usepackage{epsfig}			

\usepackage[square,comma]{natbib}

\title{Excited mesons on dynamical clover-Wilson lattices}

\ShortTitle{Excited mesons on dynamical clover-Wilson lattices}

\author{Tommy Burch, Christian Ehmann, Christian Hagen,\speaker{ Martin Hetzenegger}, Andreas Sch\"afer\\
        Institut f\"ur Theoretische Physik, Universit\"at Regensburg, D-93040 Regensburg, Germany.\\
        E-mail: \email{martin.hetzenegger@physik.uni-regensburg.de}}

\abstract{We present results for masses of excited mesons from dynamical clover-Wilson lattices provided by the CP-PACS collaboration at pion masses down to 500 MeV. Our analysis of the data is based on using a matrix of correlators from various source and sink operators. The spectroscopy results are discussed and compared to experimental values.}

\FullConference{The XXV International Symposium on Lattice Field Theory\\
		 July 30-4 August 2007\\
		 Regensburg, Germany}

\begin{document}

\section{Introduction}

We want to determine ground and excited states of light-light mesons with spin 0,1 and 2. These masses are computed in lattice simulations from the asymptotic behavior of Euclidean-time correlation functions. Our meson correlator can be written as
 	\begin{align}
	& {C(t) = \langle M(x,t) \bar{M}(0,0) \rangle}
	\end{align}
with the interpolator
	\begin{align}
	& {M(x,t) = \bar{\psi}(x,t)\hat{O}\psi(x,t)},
 	\end{align}
where the operator $\hat{O}$ represents a combination of lattice derivative operators and Dirac $\gamma$-matrices to create the desired quantum numbers.

For mesons with spin 2, we used the operators from the paper of X. Liao and T. Manke \cite{liao}.

\section{Variational Method}

To extract the masses of ground and excited states of spin-0 and spin-1 mesons we apply the variational method, which was first proposed by C. Michael \cite{Michael:1985ne} and later refined by Lüscher and Wolff \cite{Luscher:1990ck}.
The main idea is to use several different interpolators $M(x,t)_i$ , $i = 1, ...N$ with the quantum numbers of the desired states and compute all cross correlations:
	\begin{align}
	&{C(t)_{ij} = \langle M(x,t)_i \bar{M}(0,0)_j \rangle}.
	\end{align}
The  eigenvalues we obtain by solving the generalized eigenvalue problem
	\begin{align}
	&{C(t)\overrightarrow{\nu}^{(k)} = \lambda^{(k)}(t)C(t_0)\overrightarrow{\nu}^{(k)}}
	\end{align}
behave as
	\begin{align}
	&{\lambda^{(k)}(t)\propto e^{-(t-t_0)M_k}[1+O(e^{-(t-t_0)\Delta M_k})]},
	\end{align}
where $M_k$ is the mass of the $k$-th state and $\Delta M_k$ is the difference to the mass closest to $M_k$.

This method has been used previously in a quenched calculation of excited meson masses \cite{Burch:2006dg}. Preliminary dynamical results for spin 0 and 1 were reported in \cite{hagen}.

\section{Technical Details}

\begin{table}[b]
\centering
\begin{tabular}{c|c|c|c|c|c} 
	\textbf{beta} & \textbf{lattice} & $\mathbf{c_{SW}}$ & \textbf{a [fm]} & \textbf{La [fm]} & $\mathbf{\kappa}$ \textbf{values}\\ 
	\hline
	\vphantom{$X^{X^X}$}1.80 & $12^3\times24$ & 1.60 & 0.2150(22) & 2.580(26) & 0.1409, 0.1430, 0.1445, 0.1464 \\
	1.95 & $16^3\times32$ & 1.53 & 0.1555(17) & 2.489(27) & 0.1375, 0.1390, 0.1400, 0.1410 
\end{tabular}
\caption{Parameters of our calculation}
\label{tab:lattices}
\end{table}

For our calculations we use dynamical clover-Wilson lattices provided by the CP-PACS collaboration \cite{Ali Khan:2001tx},\cite{Ali Khan:2001-erratum},\cite{AokiS}  and the Chroma Software Library \cite{Edwards:2004sx},\cite{chroma1} (including BAGEL \cite{chroma2}). The parameters of the gauge configurations are collected in Table \ref{tab:lattices}.

We perform runs at four different valence and sea quark masses ($\kappa_{val}=\kappa_{sea}$), with about 100 configurations for each mass, and extrapolate to the chiral limit to obtain the physical hadron masses. In order to improve our signal we use Jacobi Smearing ($N_{iter}=16,$ width $=2.4$a), producing Gaussian-type covariant sources. In addition, we use APE Link Smearing ($N_{iter}=15, \alpha=2.5$), but only to create the source and the sink operators for the spin-2 mesons.

\section{Effective Masses}

\begin{figure}[b]
	\includegraphics[angle=270,width=7.0cm]{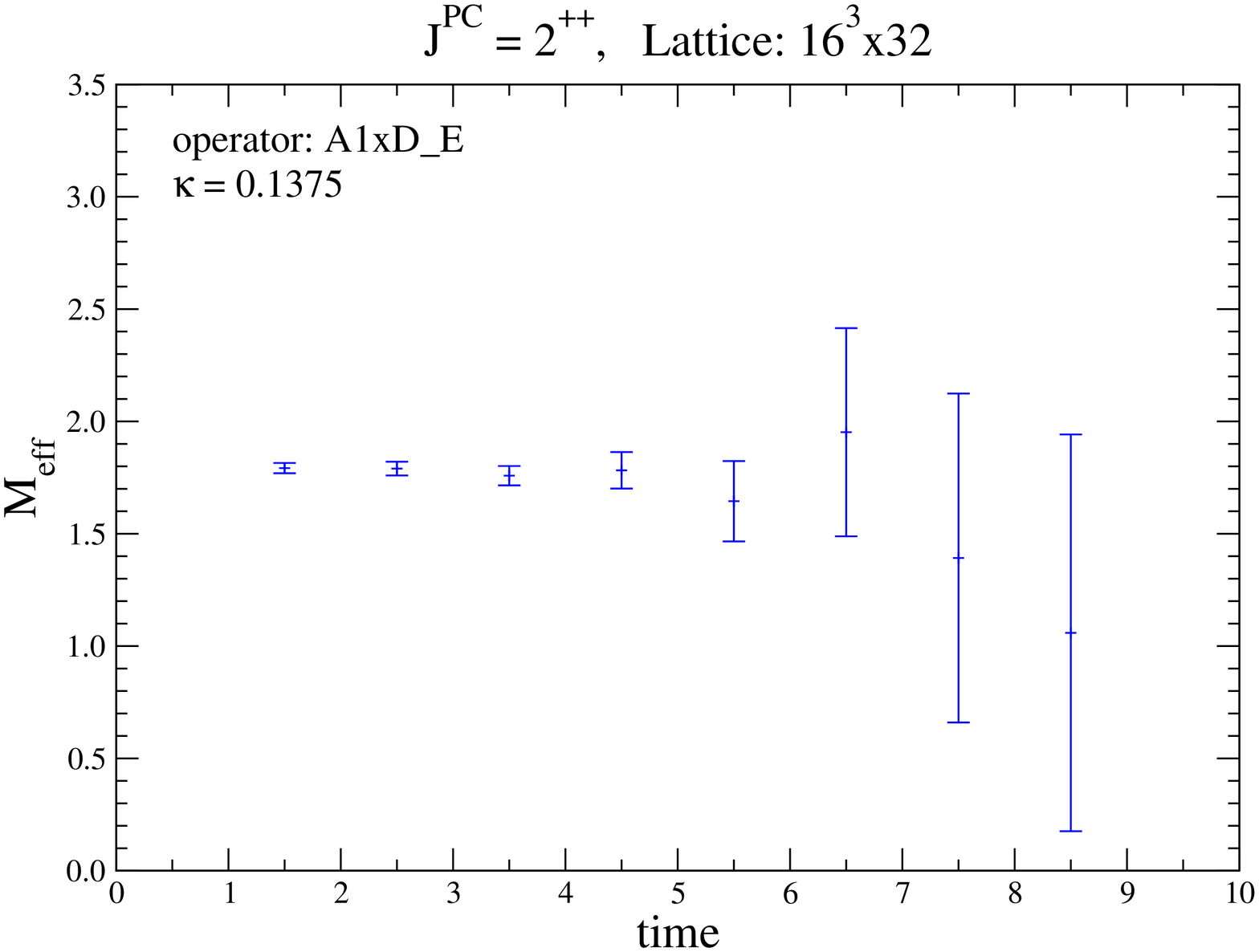} 
	\includegraphics[angle=270,width=7.0cm]{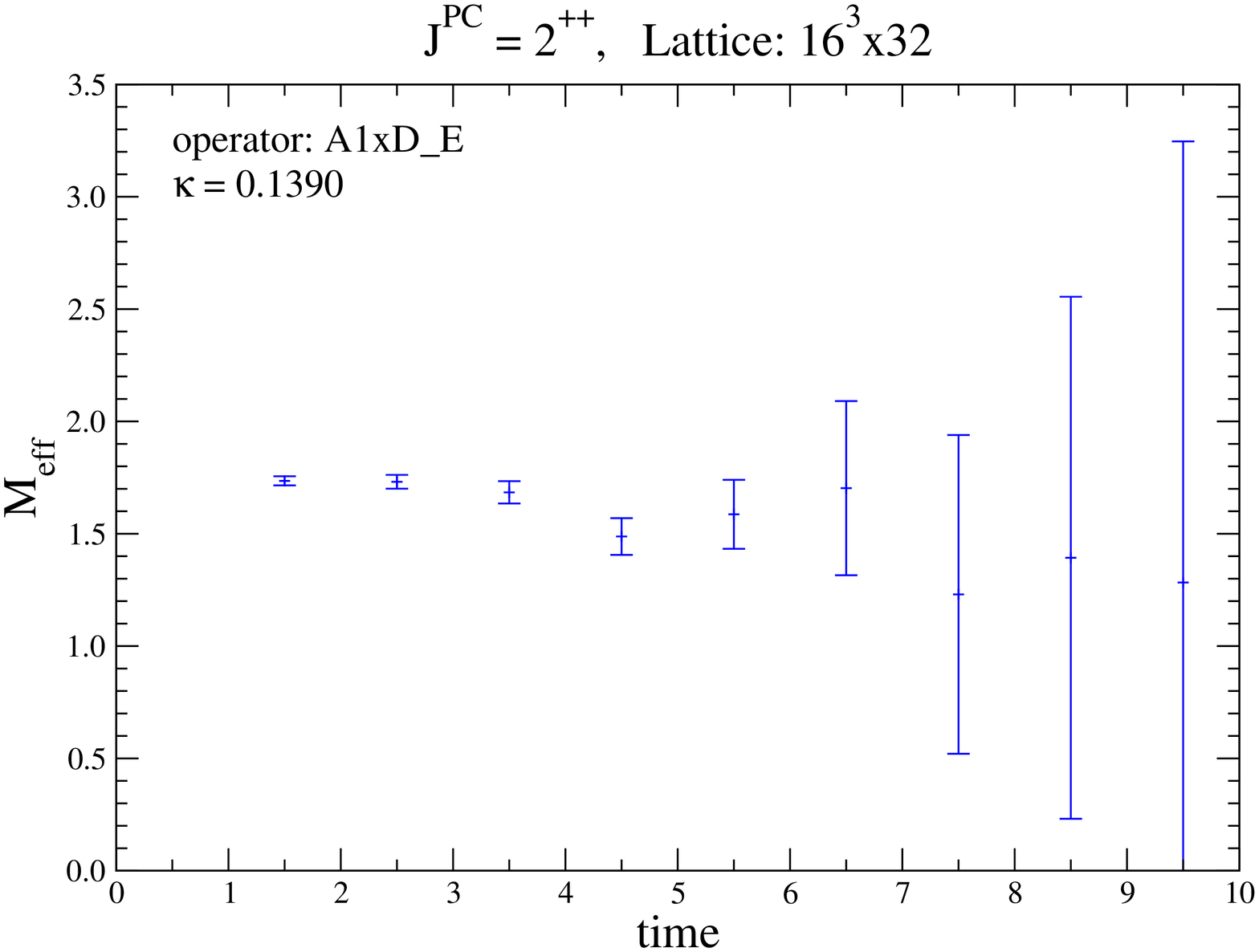} \\
	\includegraphics[angle=270,width=7.0cm]{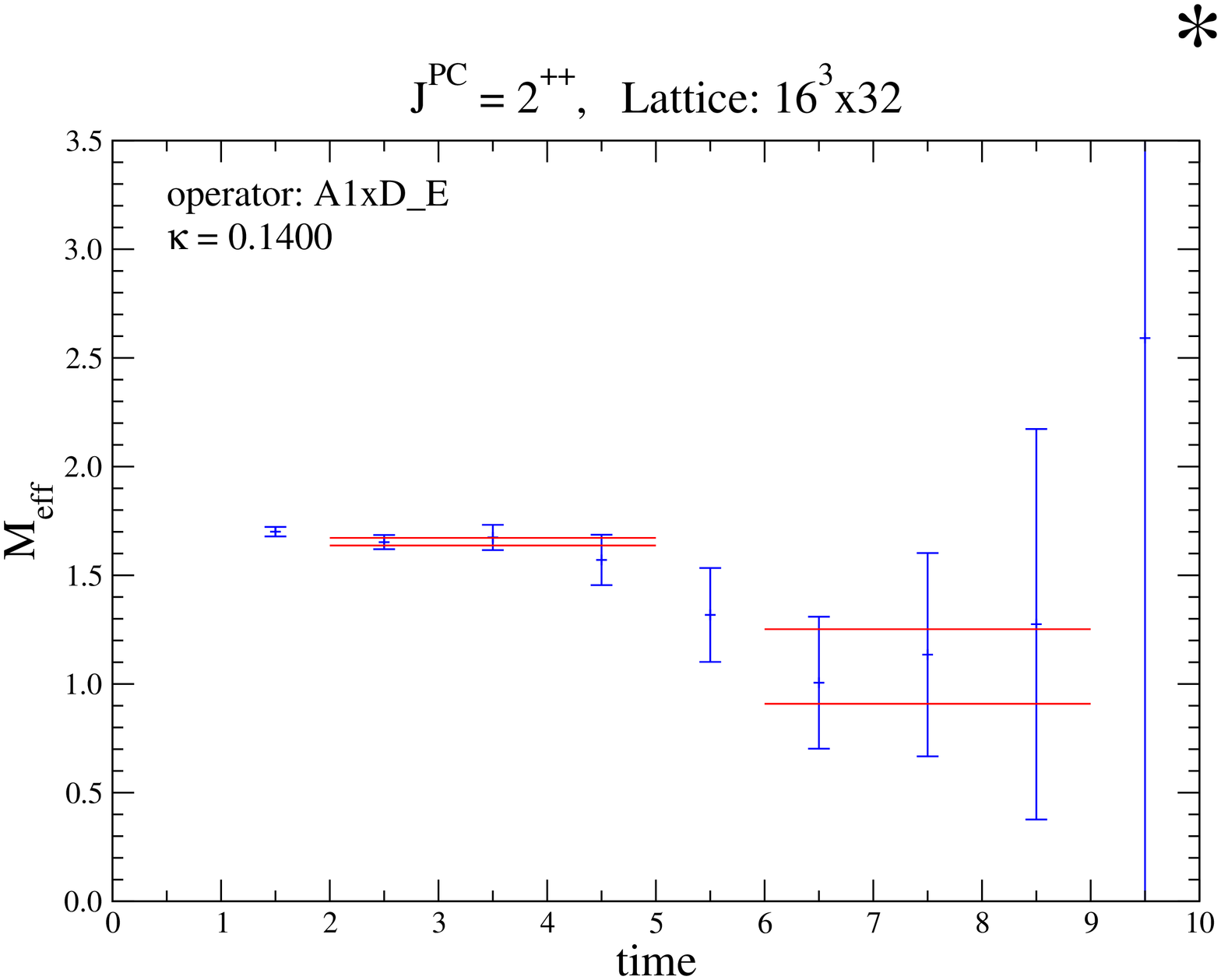} 
	\includegraphics[angle=270,width=7.0cm]{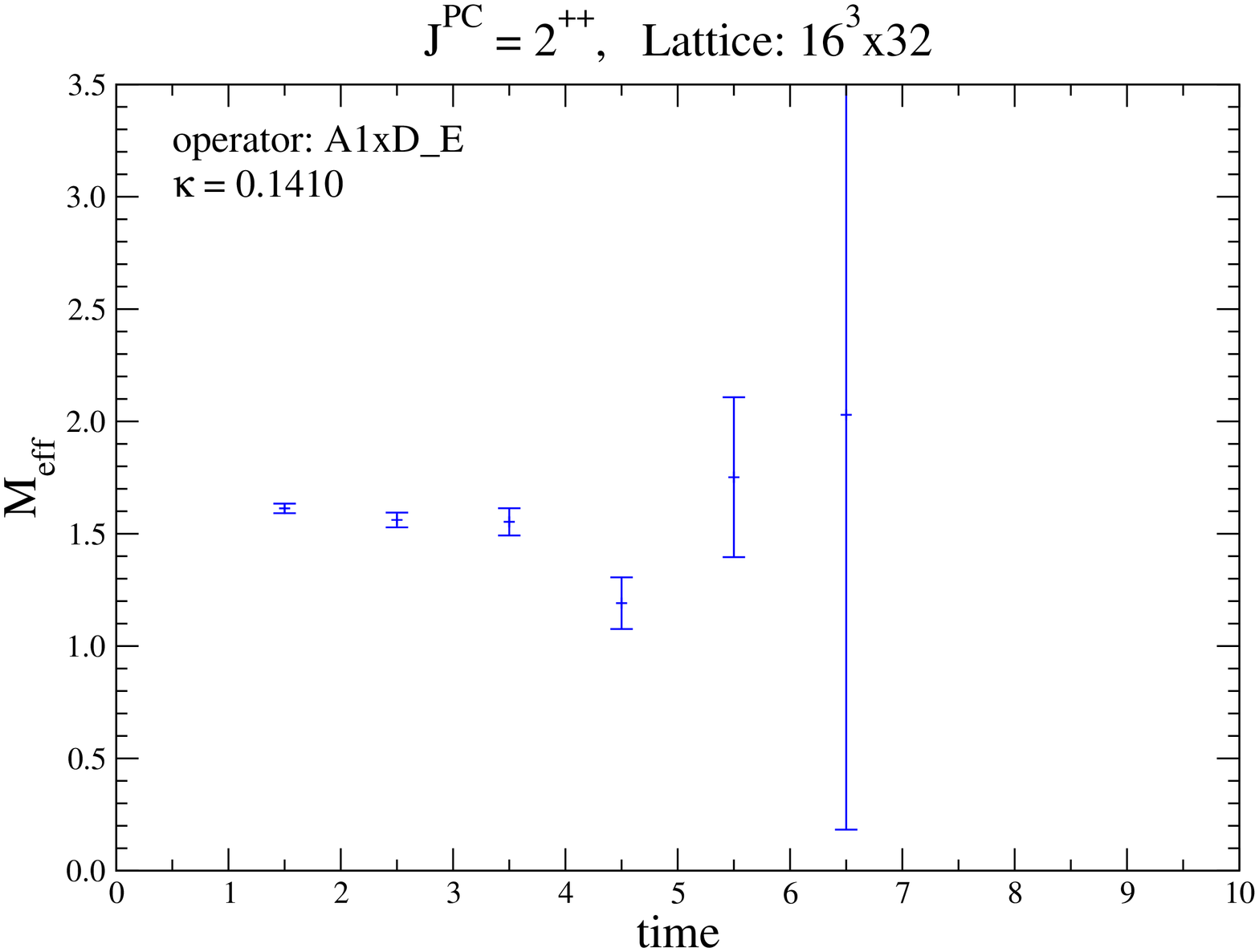} \\
\caption{Effective mass plots for a $2^{++}$ state on the $16^3\times32$ lattice.}
\label{fig:effmass}
\end{figure}
An important quantity for spectroscopy is the effective mass of a correlator. It is determined from ratios of eigenvalues on adjacent timeslices:
\begin{align}
	& {M^{(k)}_{eff}(t+\frac{1}{2}) = \ln\Bigl(\frac{\lambda^{(k)}(t)}{\lambda^{(k)}(t+1)}\Bigr)}.
\end{align}
With the help of effective mass plots, one can find appropriate fitting ranges by looking for a clear plateau. Figure \ref{fig:effmass} shows the effective masses for a meson operator that couples to a $2^{++}$ state. This state gives very good signals when compared to most of the other high-spin states. On the first plot, which shows the effective mass for the highest quark mass, a quite clear and long plateau can be seen, while the plateaus for lower quark masses become shorter and it is sometimes ambiguous where to start fitting. After all, we always try to fit at least three points. Sometimes there are even two choices for fitting ranges, as in the third plot, marked with an asterisk. In these cases we present results from different possibilities.

\section{Chiral Extrapolations}

\begin{figure}[h]
	\includegraphics[width=6.7cm,height=8cm,angle=270]{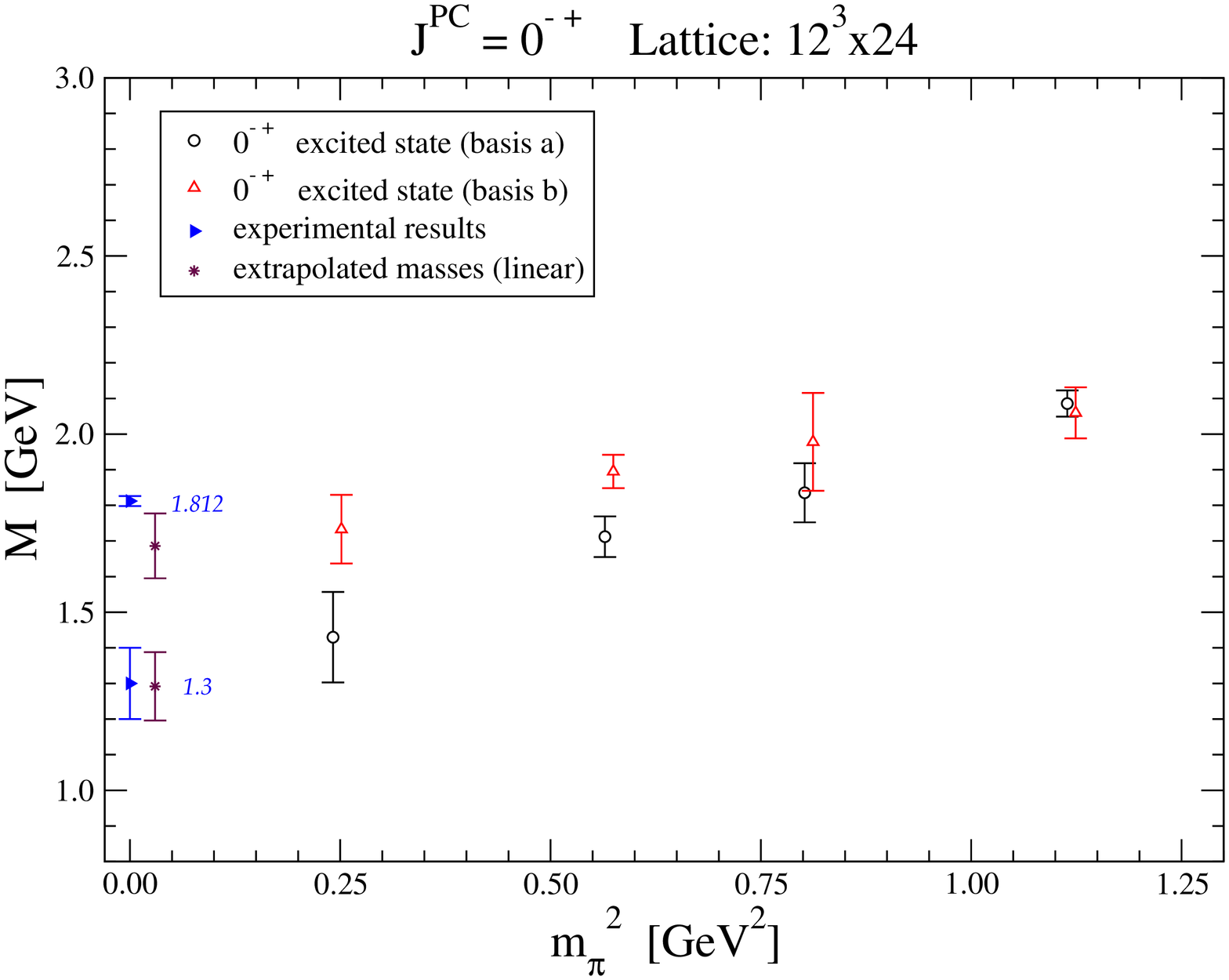} \hspace{-0.7cm}
	\includegraphics[width=6.7cm,height=8cm,angle=270]{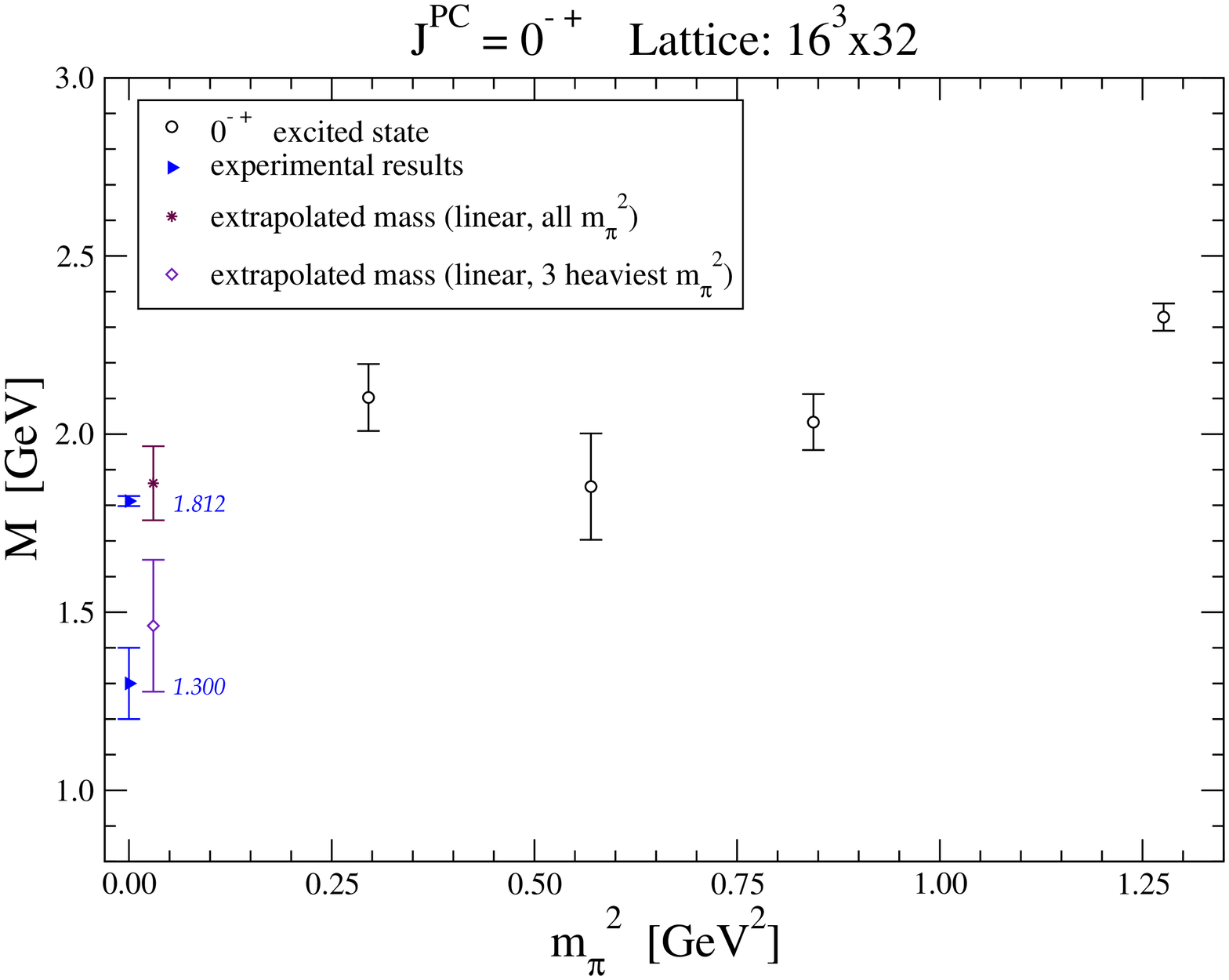} \\
	\includegraphics[width=6.7cm,height=8cm,angle=270]{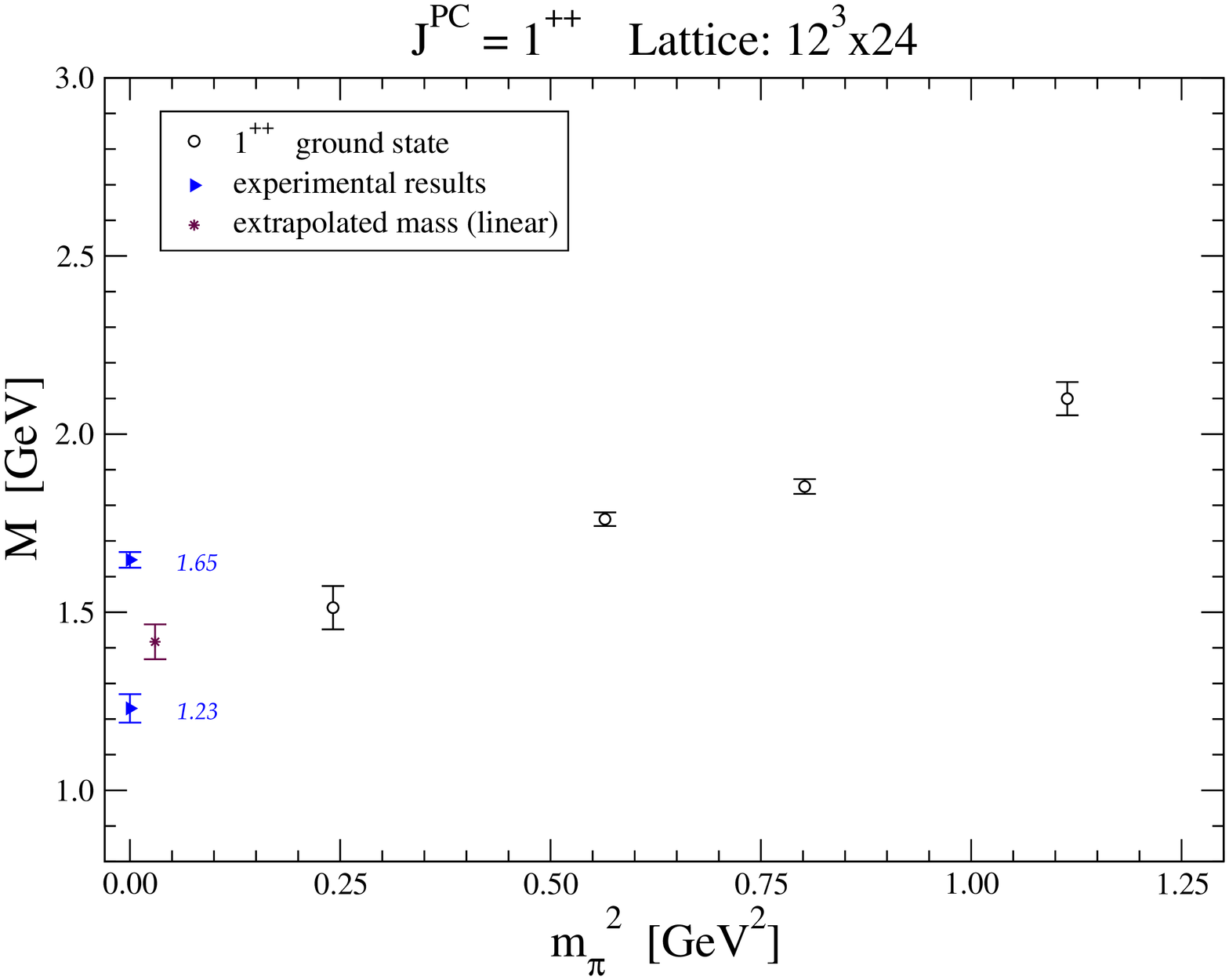} \hspace{-0.7cm}
	\includegraphics[width=6.7cm,height=8cm,angle=270]{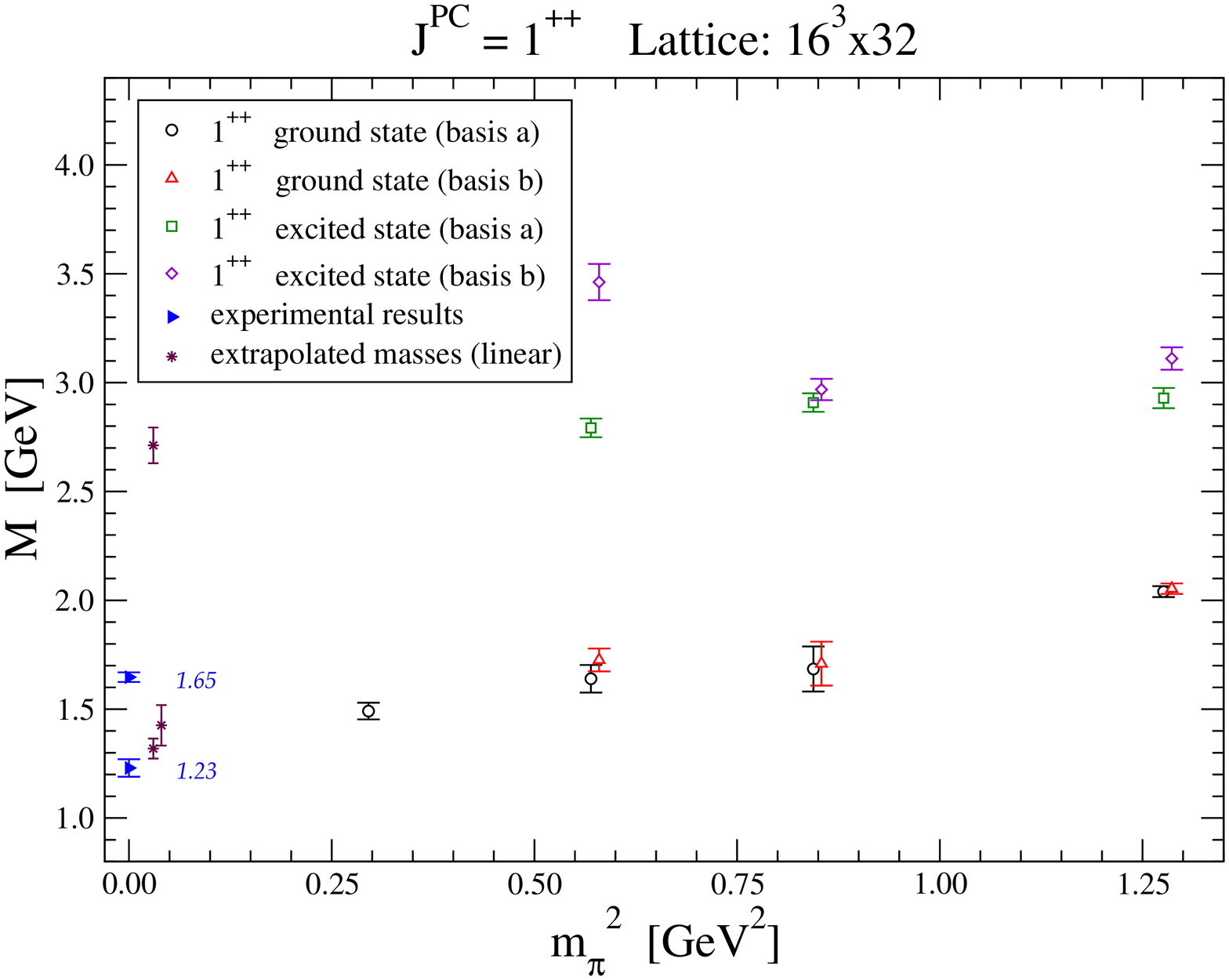} \\
\caption{Ground and excited state masses for $0^{-+}$ and $1^{++}$ mesons for both lattices. The blue triangles represent the experimental values and the purple stars mark the extrapolated masses.}
\label{fig:extrapol1}
\end{figure}
Where the data are sufficient, we perform a chiral extrapolation of our results. This is the case for mesons with quantum numbers $J^{PC} = 0^{-+}, 0^{++}, 1^{++}, 1^{--}, 2^{++}, 2^{--}$ and $2^{-+}$.

The results of these fits for the ground- and excited-state masses of the $0^{-+}$ and $1^{++}$ on the $12^3\times24$ and the $16^3\times32$ lattices are shown in Figure \ref{fig:extrapol1}. The chiral extrapolations are linear in $m^2_{\pi}$.

For some states, for example for the $1^{++}$ on the $16^3\times32$ lattice, a couple of operator combinations give the same ground and excited state (most times within errors). These are labeled as basis a, basis b and so on.

In some cases we present two extrapolations, as some points appear to be outliers. For example, for the $0^{-+}$ excited state on the $16^3\times32$ lattice, we extrapolate with all masses, as well as with only the three heaviest ones.
\begin{figure}[t]
	\includegraphics[width=6.7cm,height=8cm,angle=270]{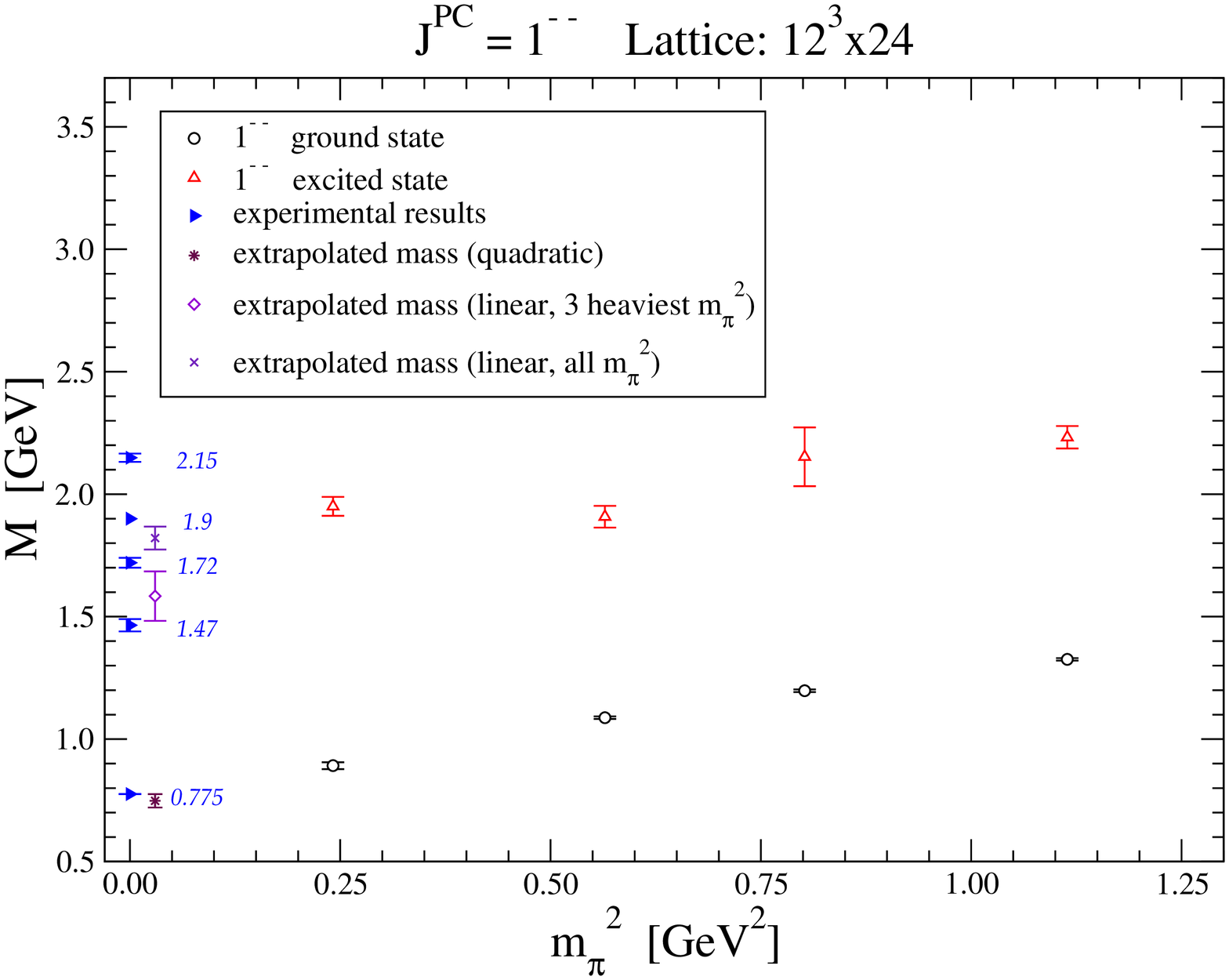}\hspace{-0.7cm}
	\includegraphics[width=6.7cm,height=8cm,angle=270]{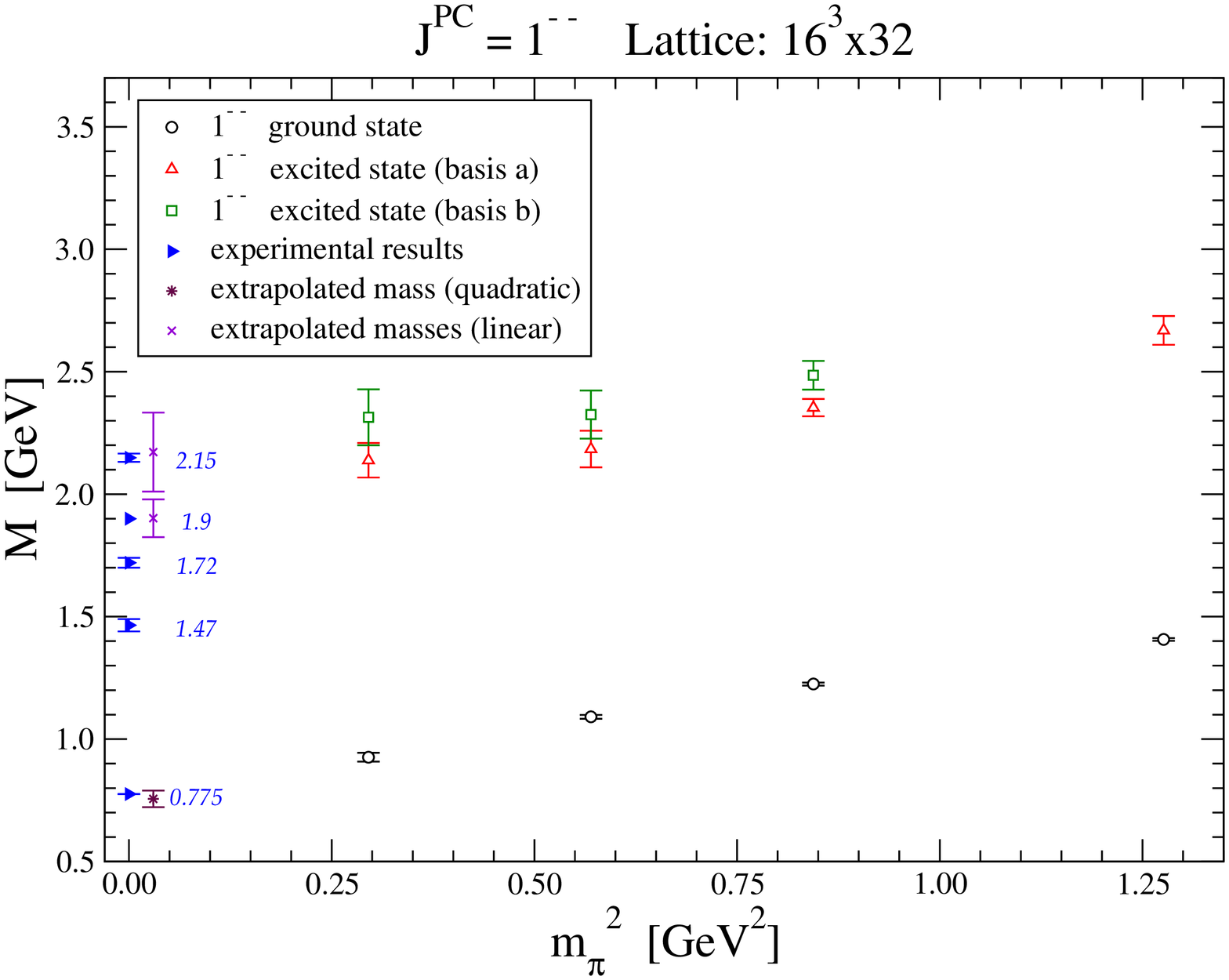}\\
	\includegraphics[width=6.7cm,height=8cm,angle=270]{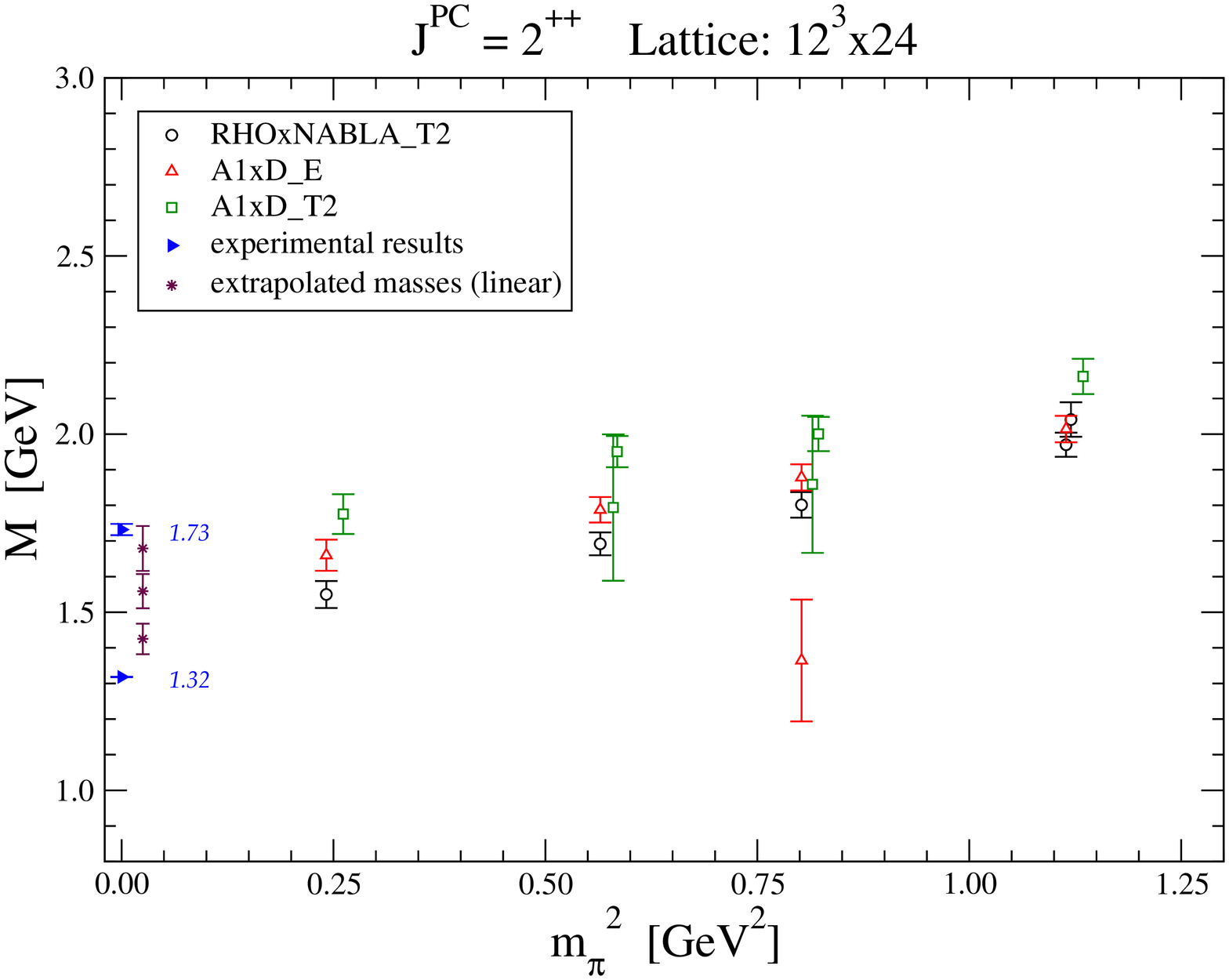}\hspace{-0.7cm}
	\includegraphics[width=6.7cm,height=8cm,angle=270]{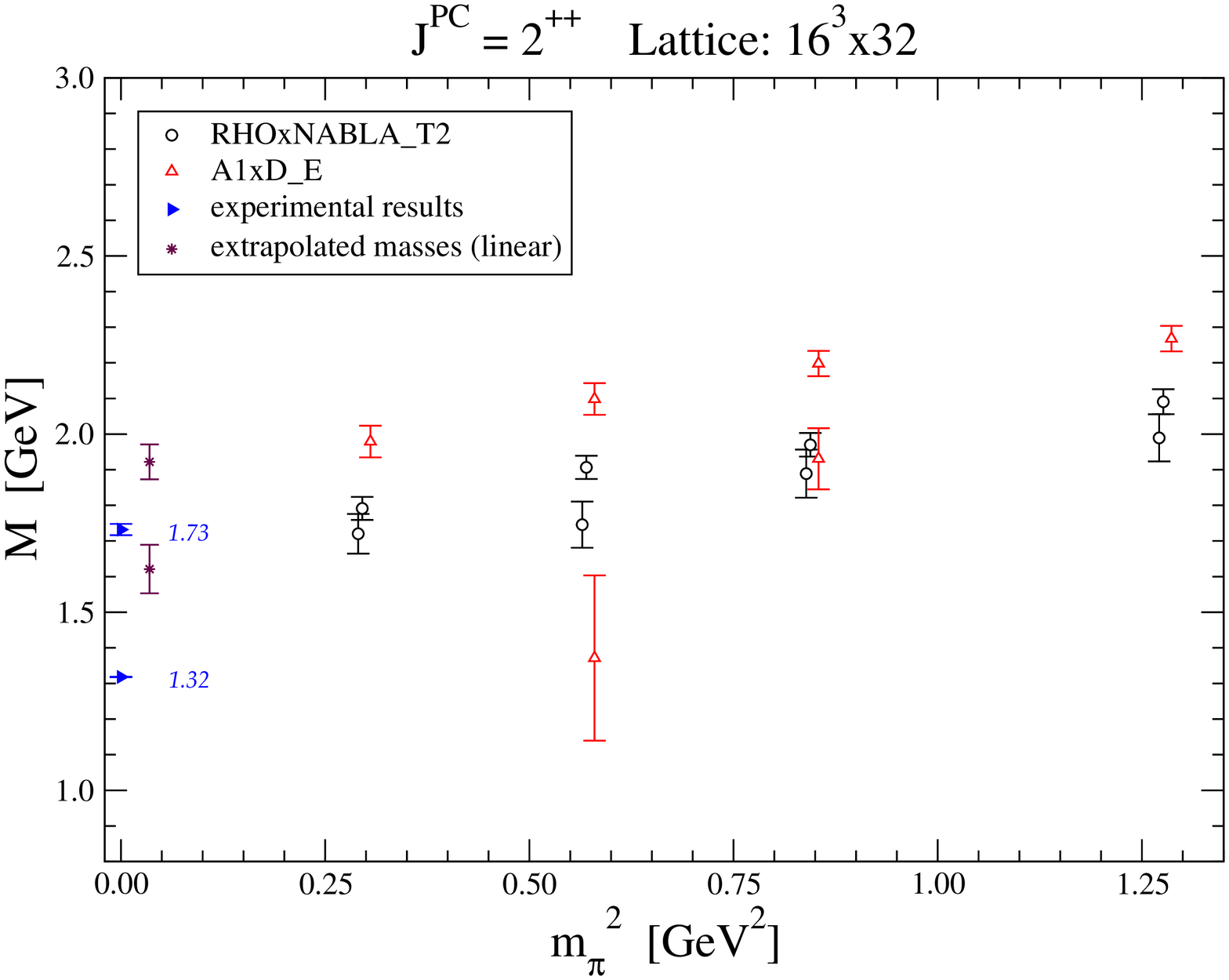}\\
\caption{Ground and excited state masses for $1^{--}$ and $2^{++}$ mesons for both lattices.}
\label{fig:extrapol2}
\end{figure}

Figure \ref{fig:extrapol2} displays the chiral extraploations of the $1^{--}$ and the $2^{++}$ states on both lattices. We perform a second order polynomial fit in $m^2_{\pi}$ for the $1^{--}$ ground state. The comparison with the experimental ground state shows good agreement.

For high-spin states we extrapolate linearly in $m^2_{\pi}$. Because we didn't calculate the cross correlations for the high-spin states yet, we do not use the variational method in these cases.

As an example we show the results for the single correlators for the $2^{++}$ state in the third and fourth plot of Figure \ref{fig:extrapol2}. They should all extrapolate to the ground state at about 1320 MeV, but our masses come out to high and don't agree within the errors. We hope that finer lattices and cross correlations will improve the situation.

\section{Conclusions \& Outlook}

\begin{figure}[Ht]
\begin{center}
\includegraphics[width=11.7cm,height=16.5cm,angle=270]{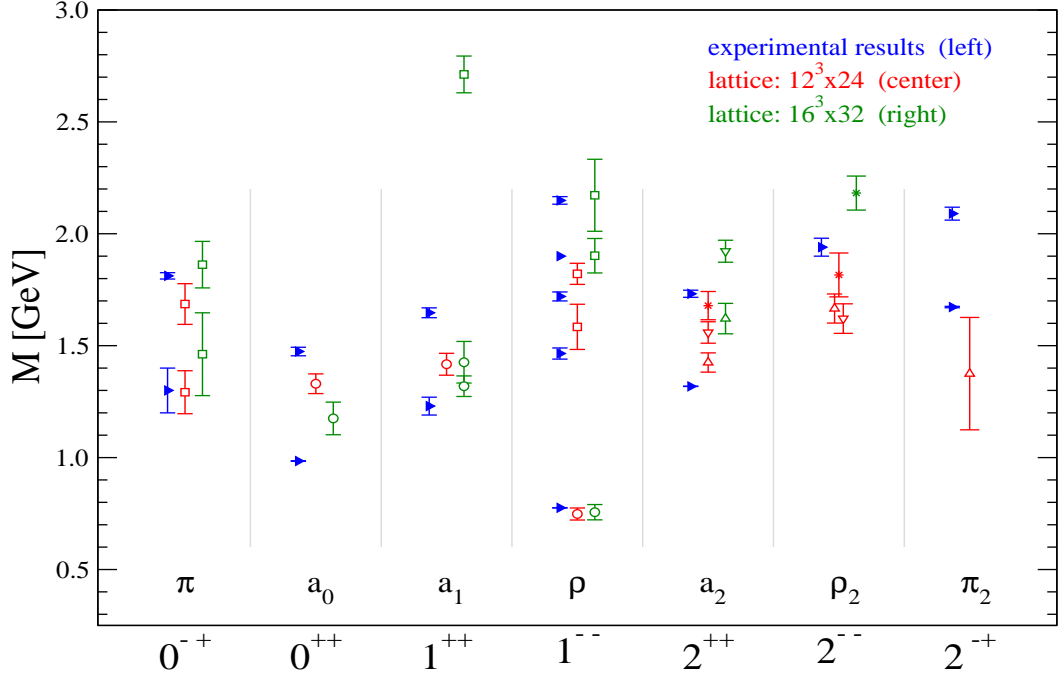}
\caption{Chiral Extrapolation of our results. For spin-0 and spin-1 states we use circles for ground states and squares for excited states. For the high-spin states different symbols represent the different operators. There are three columns for each operator. The left column is for the experimental results, the central for the $12^3\times24$ lattice and the right one for the $16^3\times32$ lattice.}
\label{fig:results}
\end{center}
\end{figure}
In this article we present a spectroscopy calculation on dynamical clover-Wilson lattices of excited and high-spin mesons using the variational method. We constructed several meson interpolators for spin 0 and 1. For the high-spin states we use the operators from the paper of X. Liao and T. Manke.

All our results are collected in Figure \ref{fig:results}. The $1^{--}$ ground state is consistent with the experimental value, but the other masses often come out too high with large error bars. This may be caused by discretization effects since these lattices are quite coarse. Also, the volume ($La\approx2.5$fm) might be too small for high-spin mesons. Furthermore, the fitting ranges in our effective masses are not always unambiguous.

For the future we plan to use cross correlations for high-spin mesons to improve the signal. We will also go to finer lattices to minimize discretization effects and perform a continuum extrapolation.

\section{Acknowledgments}

Our calculations were performed on the QCDOC in Regensburg and we thank Tilo Wettig for supplying the machine and Stefan Solbrig for his help and support. This work is supported in part by BMBF and DFG.

\end{document}